\documentclass[%
 aip,
 amsmath,amssymb,
 reprint,%
]{revtex4-1}

\usepackage{graphicx}
\usepackage{dcolumn}
\usepackage{bm}
\usepackage{tabularx}
\usepackage{algpseudocode}
\usepackage{algorithm}

\usepackage[utf8]{inputenc}
\usepackage[T1]{fontenc}
\usepackage{mathptmx}
\usepackage{etoolbox}
\usepackage{xcolor}
\usepackage{hyperref}

\makeatletter
\def\@email#1#2{%
 \endgroup
 \patchcmd{\titleblock@produce}
  {\frontmatter@RRAPformat}
  {\frontmatter@RRAPformat{\produce@RRAP{*#1\href{mailto:#2}{#2}}}\frontmatter@RRAPformat}
  {}{}
}%
\makeatother
\begin{document}


\title{Universal Convergence Metric for Time-Resolved Neutron Scattering}

\author{Chi-Huan Tung}
\affiliation{Neutron Scattering Division, Oak Ridge National Laboratory, Oak Ridge, 37831, Tennessee, United States}
\author{Lijie Ding}
\affiliation{Neutron Scattering Division, Oak Ridge National Laboratory, Oak Ridge, 37831, Tennessee, United States}
\author{Yuya Shinohara}
\affiliation{Materials Science and Technology Division, Oak Ridge National Laboratory, Oak Ridge, 37831, Tennessee, United States}
\author{Guan-Rong Huang}
\affiliation{Department of Engineering and System Science, National Tsing Hua University, Hsinchu 30013, Taiwan}
\affiliation{Physics Division, National Center for Theoretical Sciences, Taipei 10617, Taiwan}
\author{Jan-Michael Carrillo}
\affiliation{Center for Nanophase Materials Sciences, Oak Ridge National Laboratory, Oak Ridge, 37831, Tennessee, United States}
\author{Wei-Ren Chen}
\email{chenw@ornl.gov}
\affiliation{Neutron Scattering Division, Oak Ridge National Laboratory, Oak Ridge, 37831, Tennessee, United States}

\author{Changwoo Do}
\email{doc1@ornl.gov}
\affiliation{Neutron Scattering Division, Oak Ridge National Laboratory, Oak Ridge, 37831, Tennessee, United States}
\date{\today}

\begin{abstract}

This work introduces a model-independent, dimensionless metric for predicting optimal measurement duration in time-resolved Small-Angle Neutron Scattering (SANS) using early-time data. Built on a Gaussian Process Regression (GPR) framework, the method reconstructs scattering profiles with quantified uncertainty, even from sparse or noisy measurements. Demonstrated on the EQSANS instrument at the Spallation Neutron Source, the approach generalizes to general SANS instruments with a two-dimensional detector. A key result is the discovery of a dimensionless convergence metric revealing a universal power-law scaling in profile evolution across soft matter systems. When time is normalized by a system-specific characteristic time \(t^*\), the variation in inferred profiles collapses onto a single curve with an exponent between \(-2\) and \(-1\). This trend emerges within the first ten time steps, enabling early prediction of measurement sufficiency. The method supports real-time experimental optimization and is especially valuable for maximizing efficiency in low-flux environments such as compact accelerator-based neutron sources.

\end{abstract}

\maketitle

\section{Introduction}
\label{sec:introduction}

Small-Angle Neutron Scattering (SANS) is a well-established technique for probing mesoscale structures in complex condensed matter systems, with sensitivity to features in the range of 1--100~nm. It has found broad utility across materials science, soft matter physics, and biological applications, offering access to size, shape, and spatial correlations in inhomogeneous media such as polymers, colloids, and macromolecular assemblies. The unique characteristics of neutrons—their interaction with atomic nuclei and intrinsic magnetic moments—enable contrast variation, particularly through isotopic substitution of hydrogen and deuterium, and facilitate selective labeling of components within multicomponent systems.\cite{ILL1, ILL2, ILL3} Moreover, the high penetration depth of neutrons makes SANS a powerful tool for bulk and \textit{in situ} structural characterization.\cite{ILL1, ILL2, ILL3}

Despite these advantages, a central limitation of SANS remains its relatively low neutron flux compared to X-ray scattering measurements. This constraint is especially problematic when examining weakly scattering or dilute systems, where the scattered intensity is low and the signal-to-noise ratio (SNR) diminishes. Achieving statistically meaningful data under such conditions typically requires prolonged acquisition times, which can hinder the study of dynamic or time-sensitive phenomena.

This challenge is particularly acute in time-resolved SANS experiments, where the aim is to resolve structural evolution during processes such as chemical reactions, self-assembly, phase transitions, or conformational changes in biomolecules. The intrinsic timescales of these processes can be significantly shorter than the exposure durations required for high-SNR measurements, making conventional data acquisition protocols inadequate. For example, studies of nanoparticle self-assembly~\cite{forster1998amphiphilic} or enzymatic reactions may demand sub-second temporal resolution, a regime often inaccessible without substantial compromises in data quality.

While advances in source intensity and instrument design have improved throughput, such developments typically involve significant infrastructure investment. A complementary strategy for enhancing the efficiency of neutron experiments lies in the development of advanced data analysis methodologies. Traditional SANS workflows often assume static samples and rely on time-intensive data collection to achieve statistical reliability—assumptions that may fail in the context of dynamic experiments.

To address these limitations, we have previously introduced a statistical inference approach based on Gaussian Process Regression (GPR),\cite{williams1996gaussian, seeger2004gaussian, GPR, Deringer2021Gaussian} tailored to the analysis of sparse and noisy SANS data.~\cite{Tung2025} By modeling the scattering intensity as a smooth, continuous function of the momentum transfer $Q$, this method enables robust reconstruction of scattering profiles along with principled uncertainty quantification. The framework has demonstrated strong performance across simulated and experimental datasets, particularly in scenarios where conventional averaging fails to deliver reliable results.

In this work, we integrate the GPR-based inference scheme directly into the data workflow of the Extended $Q$-range Small-Angle Neutron Scattering (EQSANS) instrument at the Spallation Neutron Source (SNS). The framework supports real-time reconstruction of time-resolved scattering data, incorporating prior information and yielding uncertainty estimates that are critical for experimental interpretation. We further establish a model-independent, dimensionless convergence metric that defines sufficient measurement duration based solely on early-time data, thereby enabling predictive optimization of acquisition time. While our study focuses on EQSANS, the methodology is applicable to a broad class of SANS instruments, including those at compact accelerator-based neutron sources.

The remainder of this paper is organized as follows: Section~\ref{sec:results} introduces the mathematical formulation of the inference framework and presents validation results based on experimental data. Section~\ref{sec:conclusion} discusses the broader implications of our findings and highlights potential directions for future development.

\section{Method and Results}
\label{sec:results}

Before presenting criteria for evaluating inference performance, it is instructive to outline its mathematical foundation. Addressing the challenge of obtaining sufficient SNR efficiently in SANS, particularly for weakly scattering systems, we previously developed a statistical inference algorithm based on GPR,\cite{williams1996gaussian, GPR, Deringer2021Gaussian} demonstrating its effectiveness in SANS data analysis~\cite{Tung2025} (with implementation details in Ref.~\cite{Tung2025}). In this work, we integrate this robust GPR framework with the EQSANS instrument at SNS to enhance experimental workflows. Leveraging a Bayesian approach,~\cite{jeffreys1939, Gelman2013} GPR reconstructs the scattering intensity profile, $I(Q)$, from sparse and noisy EQSANS measurements by treating the unknown intensity as a realization of a Gaussian process. This methodology defines a probability distribution over possible intensity profiles based on a kernel function that describes the correlations between scattering intensities at different wave vectors, $Q$. Through Bayesian inference, GPR provides not only a prediction of the intensity profile but also a quantification of the associated prediction uncertainty, utilizing the posterior distribution over the intensity at unmeasured $Q$ points to estimate the profile with a defined confidence level. The inherent strength of GPR, as established in our prior work,~\cite{Tung2025} lies in its ability to yield reliable reconstructions even from limited and noisy data, a significant advantage for time-constrained experiments or the study of weakly scattering samples. The primary novelty of this contribution is the potential integration and application of this GPR-based Bayesian inference methodology within the experimental workflow of the EQSANS instrument at SNS, offering a pathway to more efficient and insightful SANS investigations.

Following a concise overview of the mathematical foundation for statistical inference applied to sparse SANS data, we now turn to an experimental case study focused on evaluating measurement sufficiency. Specifically, we demonstrate how the proposed dimensionless convergence metric can be used to assess the reliability of scattering profiles and determine the minimum acquisition time required to achieve statistically stable results.
This investigation begins by examining the spectral evolution of a D\(_2\)O solution containing cetyltrimethylammonium bromide (CTAB) and sodium salicylate (NaSal), \cite{lam2019ctab, huang2022ion} measured as a function of total exposure time using the EQSANS instrument at SNS.

\begin{figure}
    \centering
    \includegraphics[width=1\linewidth]{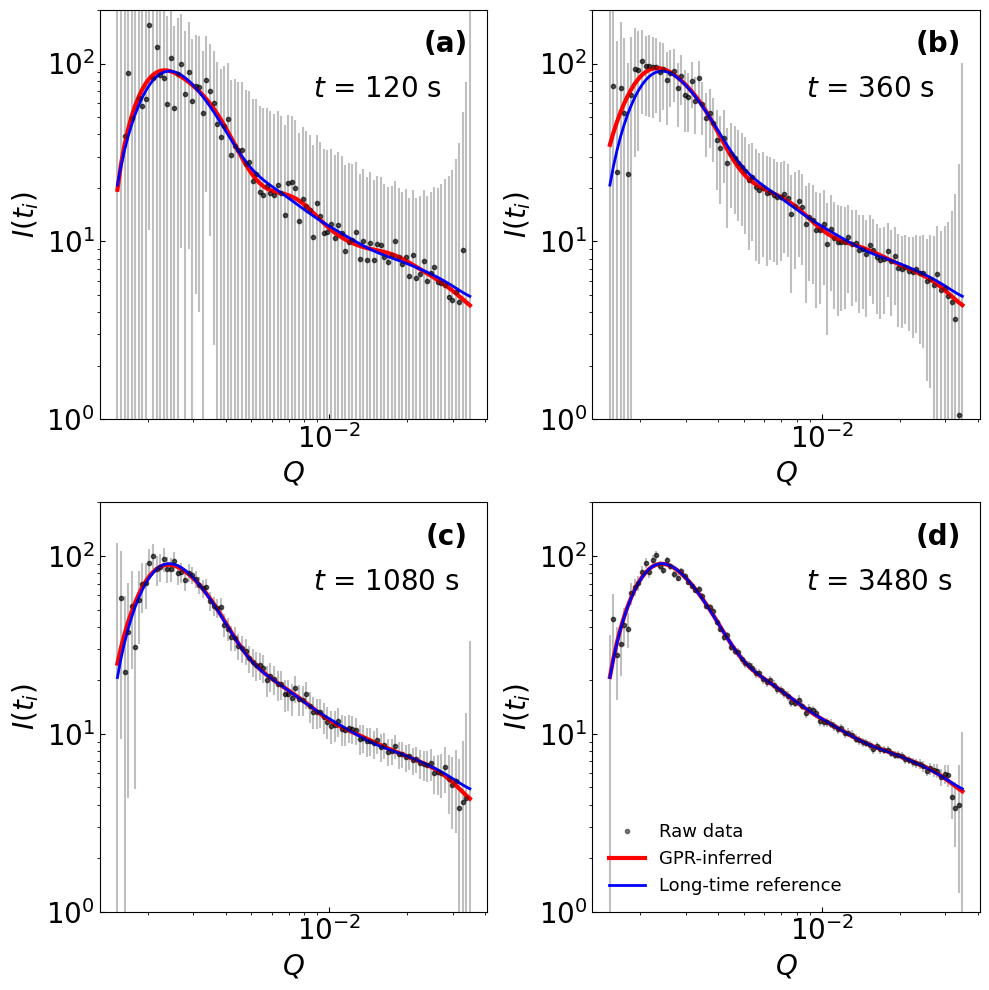}
    \caption{
    Scattering intensity \( I(Q) \) of a D\(_2\)O solution of CTAB/NaSal as a function of scattering vector magnitude \( Q \) at different cumulative exposure times: 
    (a) 120 s, (b) 360 s, (c) 1080 s, and (d) 3480 s.  Black points represent raw data with associated error bars, red curves show the GPR-denoised intensity profiles, and blue curves indicate the long-time result for reference. As exposure time increases, statistical noise in the raw data is progressively reduced, and the GPR-inferred curves converge toward the long-time result
    }
    \label{fig:1}
\end{figure}

Fig.~\ref{fig:1}(a-d) presents the experimentally measured scattering intensity \( I(Q) \) of this solution at various exposure times. The raw data exhibit inherent statistical fluctuations, with error bars indicating the uncertainty at each scattering vector magnitude \( Q \) after azimuthal averaging of two-dimensional scattering image. As expected from Poisson statistics \cite{james2006statistical}, these uncertainties decrease with longer exposure times due to improved counting statistics. Red solid lines in Fig.~\ref{fig:1} show the maximum a posteriori (MAP) estimate of scattering profiles after applying a GPR inference framework to the experimental data. The GPR reconstruction effectively suppresses noise while preserving the essential features of the scattering profile. It should be noted that the uncertainty of the MAP estimate can also be quantified within the GPR framework. Under the assumption of uniform sampling in \( Q \), the uncertainty exhibits a similar \( Q \)-dependence to that of the raw experimental data. A more detailed discussion on the estimation and interpretation of uncertainty within the probabilistic inference framework is provided in our previous work.~\cite{Tung2025} As the data collection time increases in Fig.~\ref{fig:1}, the inferred intensity curves converge towards the long-exposure reference, indicating reduced measurement variability with increased acquisition time. This trend suggests a practical criterion for determining sufficient measurement duration: the point at which additional exposure yields negligible change in the inferred profile.

\begin{figure*}
    \centering
    \includegraphics[width=1\linewidth]{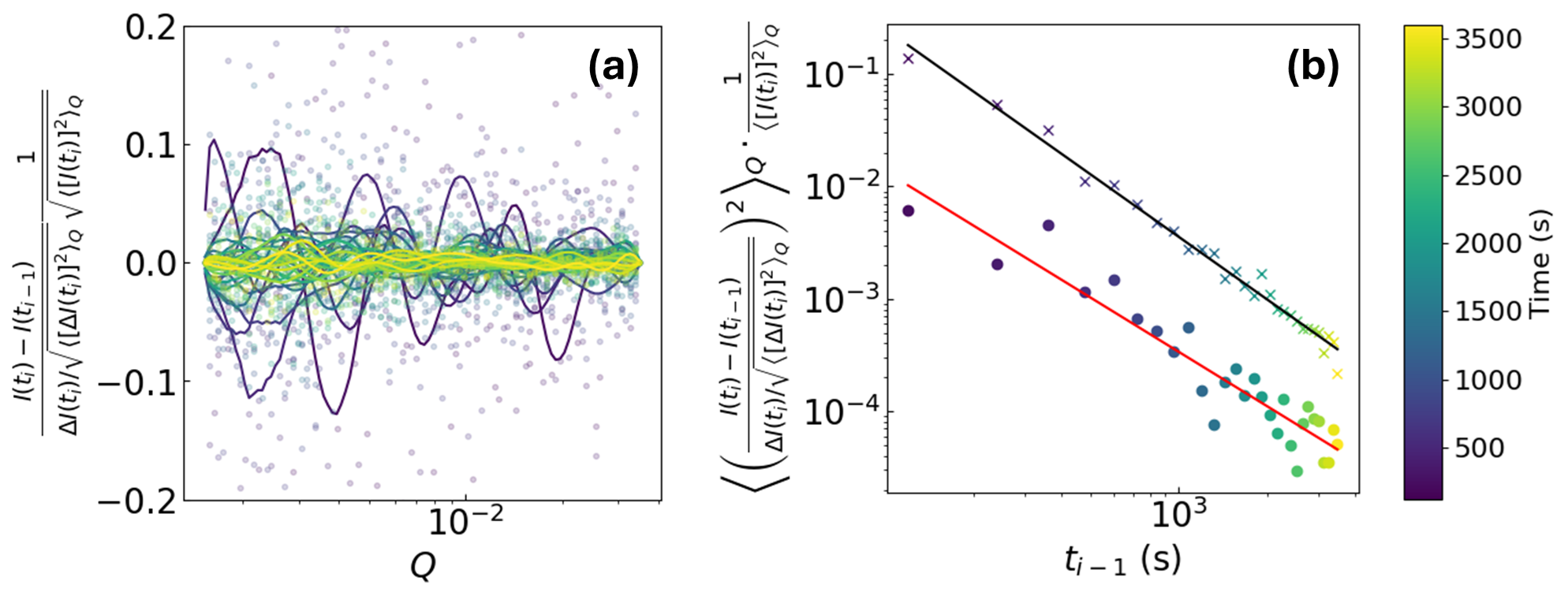}
    \caption{(a) Dimensionless relative change computed using Eqn.~\ref{eq:relative_change} for both raw experimental data (markers) and GPR-inferred scattering profiles (curves). (b) Mean relative variation defined in Eqn.~\ref{eq:mean_relative_variation} plotted against total measurement time. A power-law scaling trend is observed for both raw and inferred data, consistent with statistical expectations.}
    \label{fig:2}
\end{figure*}
In order to derive a robust criterion for sufficient measurement duration, we evaluate the cumulative scattering intensity profile at time \( t_i \) in comparison with that obtained at the previous time step \( t_{i-1} \). The measurement is deemed sufficient once the change in intensity, \( I(t_i) - I(t_{i-1}) \), falls below an appropriate threshold. To account for the varying uncertainty across the scattering vector \( Q \), the intensity difference is normalized by a dimensionless factor \( \Delta I(t_i) / \sqrt{ \left\langle \left[ \Delta I(t_i) \right]^2 \right\rangle_Q } \), where \( \Delta I(t_i) \) is the uncertainty in intensity from the experiment and \( \left\langle \cdots \right\rangle_Q \) denotes averaging over \( Q \). Furthermore, to enable comparisons across different datasets, the result is scaled by the square root of the average squared intensity, \( \sqrt{ \left\langle \left[ I(t_i) \right]^2 \right\rangle_Q } \). Consequently, the following dimensionless quantity is used to characterize the relative change between two adjacent measurement times:
\begin{equation}
\label{eq:relative_change}
\frac{ I(t_i) - I(t_{i-1}) }{ \Delta I(t_i) / \sqrt{ \left\langle \left[ \Delta I(t_i) \right]^2 \right\rangle_Q } } \cdot \frac{1}{ \sqrt{ \left\langle \left[ I(t_i) \right]^2 \right\rangle_Q } },
\end{equation}
serving as an indicator of the degree to which additional measurement improves the statistical stability of the inferred scattering profile. This quantity does not require knowledge of the ground truth or the long-time limit of \( I(Q) \), and is directly computable from the measured data alone. Moreover, the formulation is broadly applicable to any scattering intensity profile, making it a general and model-independent indicator of measurement sufficiency.

Fig.~\ref{fig:2}(a) presents the evaluation of the dimensionless quantity defined in Eqn.~\ref{eq:relative_change}, computed from both the raw experimental data (shown as markers) and the probabilistic inference results obtained via GPR (shown as curves). As the measurement time increases, the relative variation between successive intensity profiles visibly decreases, indicating improved statistical stability in the collected data. To quantitatively capture this trend, we evaluate the \( Q \)-averaged mean relative variation:
\begin{equation}
\label{eq:mean_relative_variation}
\left\langle \left( \frac{ I(t_i) - I(t_{i-1}) }{ \Delta I(t_i) / \sqrt{ \left\langle \left[ \Delta I(t_i) \right]^2 \right\rangle_Q } } \right)^2 \right\rangle_Q \cdot \frac{1}{ \left\langle \left[ I(t_i) \right]^2 \right\rangle_Q },
\end{equation}
and plot its evolution as a function of total measurement time in Fig.~\ref{fig:2}(b). The result reveals a clear power-law relationship between measurement duration and variation magnitude, evidenced by the approximately linear trend observed on the log--log scale. Notably, the mean relative variation computed from the GPR-inferred profiles is consistently about an order of magnitude smaller than that obtained from the raw data, indicating that statistical convergence is achieved much earlier when inference is applied. This empirical scaling behavior further supports the use of Eqn.~\ref{eq:relative_change} as a meaningful indicator of measurement sufficiency and convergence in time-resolved scattering experiments.

\begin{figure*}
    \centering
    \includegraphics[width=1\linewidth]{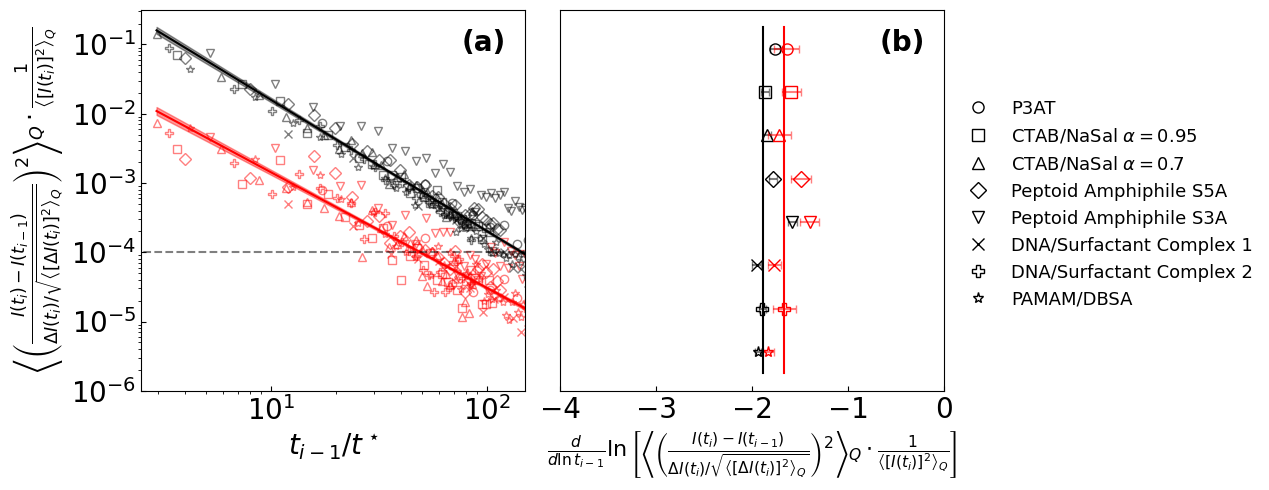}
    \caption{(a) Normalized mean relative variation across different soft matter systems, rescaled by system-specific characteristic time \( t^\star \). A collapse to a universal power-law trend is observed. The grey dashed line marks the 1\% variation level relative to the signal magnitude. (b) Extracted scaling exponents for each system individually. The vertical lines show the global best-fit exponent for both results and are around -2.}
    \label{fig:3}
\end{figure*}

To further assess the generality of the measurement sufficiency criterion, we analyzed the behavior of the mean relative variation across a range of soft matter systems. These included: a solution of conjugated polymers with poly(3-alkylthiophene) (P3AT) backbones and alkyl side chains; an aqueous solution of cetyltrimethylammonium bromide (CTAB) in 0.95 M sodium salicylate (NaSal) with $\alpha = 1$; an aqueous CTAB/NaSal solution with $\alpha = 0.7$, where $\alpha$ denotes the atomic fraction of $\mathrm{D_2O}$ in the mixed solvent system ($\mathrm{H_2O} + \mathrm{D_2O}$); a $\mathrm{D_2O}$ solution of a peptoid amphiphile; a $\mathrm{D_2O}$ solution of a DNA complex with a pH-sensitive gemini surfactant; and a $\mathrm{D_2O}$ solution of a self-assembled PAMAM dendrimer with dodecylbenzene sulfonic acid (DBSA). Fig.~\ref{fig:3}(a) shows the normalized results, where the measurement time is rescaled by a system-specific characteristic time \( t^\star \), defined such that the mean relative variation for the raw data equals 1 at \( t = t^\star \). This normalization aligns the data from different systems at a common statistical reference point, corresponding to the time when the variation between successive measurements is on the order of the signal variance itself. Remarkably, both the raw data and GPR-inferred results collapse onto distinct and consistent linear trends in the log--log plot, indicating a power-law dependence. The shaded regions represent the confidence intervals obtained from linear regression fits to the logarithmic data.

Fig.~\ref{fig:3}(b) shows the extracted scaling exponents for each system individually, with the solid line representing the overall best-fit trend across all systems. It is clear that the extracted scaling exponents cluster around \(-2\), suggesting that the \(l\)-th cumulant of the scattering profiles scales as \(O(n^{1-l})\), where \(n\) denotes the neutron count at a detector pixel and \(l \geq 2\). Consequently, during long-duration measurements, higher-order cumulants (\(l \geq 3\)) diminish quickly, and the scattering intensity converges to a normal distribution characterized solely by the first two cumulants: the mean and the variance. this behavior has a well-defined statistical interpretation. In standard neutron data reduction workflow, the measurement uncertainty for each detector pixel is estimated using Poisson counting statistics, where the standard deviation of \( n \) is \( \Delta n \propto \sqrt{n} \) and \( \Delta I \propto \sqrt{n}/n = 1/\sqrt{n} \). However, at very short exposure times, a substantial fraction of detector pixels receive zero neutron counts. To prevent division-by-zero errors or undefined values in downstream processing, these zero-count bins are typically reassigned to have errors of a value of one, corresponding to the lowest sensitivity limit of the instrument.

As a result, in this extremely low-count regime, the uncertainty \( \Delta I \) scales as
\begin{equation}
\label{eq:low_count_uncertainty}
\Delta I \propto \frac{1}{n} \propto \frac{1}{t},
\end{equation}
assuming a constant neutron flux such that the expected count \( n \propto t \). Now, consider two successive measurements \( I(t_i) \) and \( I(t_{i-1}) \), modeled as independent random variables governed by Poisson statistics. Their difference follows a Skellam distribution, and its variance can be approximated by
\begin{equation}
\label{eq:diff_variance_low}
\mathrm{Var}[I(t_i) - I(t_{i-1})] = \mathrm{Var}[I(t_i)] + \mathrm{Var}[I(t_{i-1})] \propto \frac{1}{t_{i}^2} + \frac{1}{t_{i-1}^2}.
\end{equation}
However, the last proportionality will be dominated by the earlier time, 
\begin{equation}
\label{eq:time}
\frac{1}{t_{i}^2} + \frac{1}{t_{i-1}^2} \approx \frac{2}{t_{i-1}^2}. 
\end{equation}

The numerator of Eqn.~\ref{eq:mean_relative_variation} thus scales as \( 1/t_{i-1} \), and the normalization factor \\
\( \Delta I(t_i)/\sqrt{\left\langle [\Delta I(t_i)]^2 \right\rangle_Q} \) is approximately constant since both numerator and denominator scale as \( 1/t_{i-1} \). Therefore, the squared ratio scales as \( 1/t_{i-1}^2 \), and the average intensity \( \left\langle I^2(t_i) \right\rangle_Q \) is approximately constant in this regime. Altogether, the full expression scales as:

\begin{equation}
\label{eq:tinv2_scaling}
\left\langle \left( \frac{ I(t_i) - I(t_{i-1}) }{ \Delta I(t_i)/\sqrt{ \left\langle \left[ \Delta I(t_i) \right]^2 \right\rangle_Q } } \right)^2 \right\rangle_Q \cdot \frac{1}{ \left\langle I^2(t_i) \right\rangle_Q }
\propto \frac{(t_{i-1}^{-1})^2}{1}\cdot\frac{1}{1} = t_{i-1}^{-2}.
\end{equation}

On the other hand, at sufficiently long exposure times, the neutron count becomes large enough that nearly all detector pixels register nonzero counts. In this regime, the uncertainty returns to the standard Poisson scaling,
\begin{equation}
\label{eq:high_count_uncertainty}
\Delta I \propto \frac{1}{\sqrt{t_{i-1}}}.
\end{equation}
The variance of the difference \( I(t_i) - I(t_{i-1}) \) therefore scales as \( 1/t_{i-1} \). Therefore, the full expression in Eqn.~\ref{eq:mean_relative_variation} now scales as:
\begin{equation}
\label{eq:tinv_scaling}
\left\langle \left( \frac{ I(t_i) - I(t_{i-1}) }{ \Delta I(t_i)/\sqrt{ \left\langle \left[ \Delta I(t_i) \right]^2 \right\rangle_Q } } \right)^2 \right\rangle_Q \cdot \frac{1}{ \left\langle I^2(t_i) \right\rangle_Q }
\propto \frac{(t_{i-1}^{-1/2})^2}{1}\cdot\frac{1}{1} = t_{i-1}^{-1}.
\end{equation}

This \( t^{-1} \) scaling reflects the behavior predicted by the Central Limit Theorem (CLT),\cite{huang1991statistical, kardar2007statistical} as the difference between two independent intensity estimates converges to a normal distribution whose variance decreases inversely with the effective sample size, which is proportional to measurement time. Consequently, the observed slope in the log--log plot of normalized mean relative variation versus time is expected to lie between \(-2\) and \(-1\). It is important to note that empty bins are commonly encountered in 2D scattering experiments due to the combined effects of \( Q \)-space resolution and strong spatial variation in \( I(Q) \), resulting in exponent close to \(-2\). Regardless of this exponent, the 1\% variation level relative to the signal magnitude as shown as dashed line in Fig.~\ref{fig:3} (a) already reflect statistically sufficient measurement for most cases. Therefore, the observed power-law and the normalized mean relative variation in Eqn.~\eqref{eq:mean_relative_variation} provide a statistically meaningful and experimentally accessible criterion for determining sufficient measurement time, especially at the early stages of data acquisition.

\begin{figure}
    \centering
    \includegraphics[width=1\linewidth]{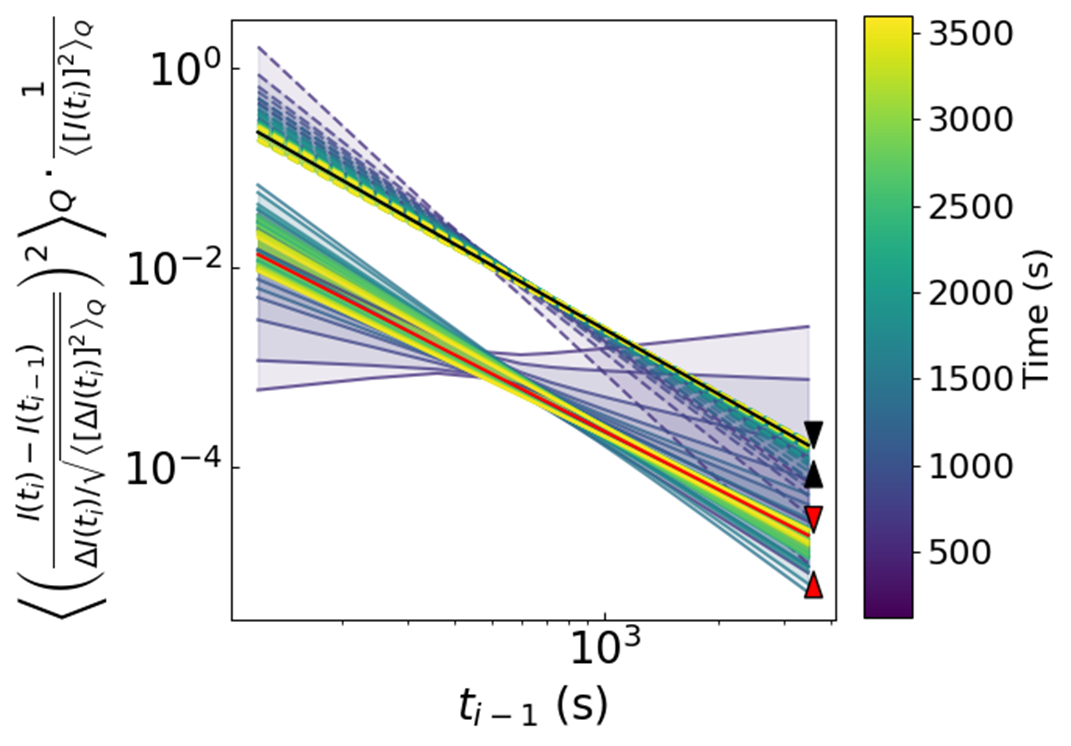}
    \caption{
    Confidence intervals derived from linear regression of the logarithm of measurement time versus the logarithm of the normalized mean relative variation (Eqn.~\ref{eq:mean_relative_variation}) for the same CTAB/NaSal dataset shown in Fig.~\ref{fig:1} and Fig.~\ref{fig:2}. Dashed lines correspond to fits using raw experimental data, while solid lines represent GPR-denoised results. Shaded regions indicate the associated 1 $\sigma$ confidence intervals. Arrows indicate the predicted uncertainty bounds at $t = 3600$~s for raw data (black) and GPR-inferred results (red), respectively. Forecasts based on early-time data show convergence within the first ten measurement steps, supporting the use of this metric as a predictive criterion for sufficient measurement duration. 
    }
    \label{fig:4}
\end{figure}

Figure~\ref{fig:4} displays the \(1\sigma\) confidence intervals obtained from linear regression between the logarithm of measurement time and the logarithm of the normalized mean relative variation, as defined in Eqn.~\ref{eq:mean_relative_variation}. Dashed lines represent fits to the raw experimental data, while solid curves correspond to the results obtained after GPR-based denoising. The shaded regions indicate the statistical uncertainty of the fitted trend. In addition, arrows mark the forecasted confidence intervals of the normalized mean relative variation at later time points, based on regression results accumulated up to earlier timesteps.
One can clearly observe that after collecting only the first ten time steps, the regression lines converge to a narrow envelope, and the uncertainty in predicted normalized mean relative variation falls below a level of approximately 2. This rapid stabilization of regression result supports the practical feasibility of using the normalized mean relative variation as a forecasting criterion for determining sufficient measurement duration.

\section{Conclusions}
\label{sec:conclusion}

In this work, we developed and implemented a robust statistical inference framework based on Gaussian Process Regression (GPR) to enhance the analysis of sparse Small-Angle Neutron Scattering (SANS) data. Integrated into the EQSANS instrument at the Spallation Neutron Source (SNS), the framework reliably reconstructs scattering profiles under low signal-to-noise conditions and limited counting statistics. It provides improved temporal resolution and quantified uncertainty, addressing key challenges in time-constrained neutron experiments.

A central contribution is the discovery of a universal power-law scaling that governs the convergence of inferred profiles with measurement time. By normalizing time with a system-specific characteristic scale \(t^\ast\), we show that the relative variation in intensity profiles collapses onto a single curve across diverse soft matter systems, with a consistent exponent between \(-2\) to \(-1\) depending on the characteristics of scattering features and statistics. This model-independent trend establishes a general, experimentally accessible criterion for determining sufficient measurement duration, broadly applicable to SANS instruments using two-dimensional detectors.

Building on this insight, we introduce a forecasting strategy based on early-time regression of the normalized variation metric. As illustrated in Fig.~\ref{fig:4}, convergence stabilizes rapidly—typically within the first ten time steps—enabling predictive termination of experiments before full data acquisition concludes. This capability supports adaptive, real-time optimization of experiment duration, minimizing exposure while ensuring statistical confidence.

The broader impact of this approach extends to a wide range of neutron and X-ray scattering platforms. It is particularly valuable for compact accelerator-based neutron sources (CANS) \cite{carpenter2016research, Bruckel2020, IAEATECDOC1981, ott2023icone}, where low flux demands high measurement efficiency. The method also applies to laboratory-based SAXS instruments and time-resolved studies of weakly scattering or radiation-sensitive systems. By enabling reliable inference from fewer detected events, the GPR framework facilitates access to kinetic processes and enhances beamline utilization.

Future directions include extending the methodology to two-dimensional and anisotropic scattering, incorporating prior structural knowledge, and developing real-time feedback systems for on-the-fly experimental control. We are also building user-friendly software tools to promote broader adoption across the scattering community.

\begin{acknowledgments}
We extend our sincere gratitude to Sidney Yip, Gilbert Strang, James S. Langer, Thomas Gutberlet, and Chun-Keung Loong for their insightful communications. This research at ORNL's Spallation Neutron Source was sponsored by the Scientific User Facilities Division, Office of Basic Energy Sciences, U.S. Department of Energy. This research was also supported by the Laboratory Directed Research and Development Program of Oak Ridge National Laboratory, managed by UT-Battelle, LLC, for the U.S. Department of Energy. Beam time was allocated to EQSANS under proposal numbers IPTS-22170.1, 22386.1, 23463.1 and 25953.1. Y.S. was supported by the U.S. Department of Energy, Office of Science, Office of Basic Energy Sciences, Materials Sciences and Engineering Division, under Contract No. DE-AC05-00OR22725. G.R.H. is supported by the National Science and Technology Council (NSTC) in Taiwan with Grant No. NSTC 111-2112-M-110-021-MY3 and NSTC 113-2112-M-029-007. A portion of this research was performed at the Center for Nanophase Materials Sciences, which is a DOE Office of Science User Facilities operated by Oak Ridge National Laboratory. 
\end{acknowledgments}

\section*{Data Availability Statement}
The data that support the findings of this study are available from the corresponding author upon reasonable request.


\begin{thebibliography}{20}%
\makeatletter
\providecommand \@ifxundefined [1]{%
 \@ifx{#1\undefined}
}%
\providecommand \@ifnum [1]{%
 \ifnum #1\expandafter \@firstoftwo
 \else \expandafter \@secondoftwo
 \fi
}%
\providecommand \@ifx [1]{%
 \ifx #1\expandafter \@firstoftwo
 \else \expandafter \@secondoftwo
 \fi
}%
\providecommand \natexlab [1]{#1}%
\providecommand \enquote  [1]{``#1''}%
\providecommand \bibnamefont  [1]{#1}%
\providecommand \bibfnamefont [1]{#1}%
\providecommand \citenamefont [1]{#1}%
\providecommand \href@noop [0]{\@secondoftwo}%
\providecommand \href [0]{\begingroup \@sanitize@url \@href}%
\providecommand \@href[1]{\@@startlink{#1}\@@href}%
\providecommand \@@href[1]{\endgroup#1\@@endlink}%
\providecommand \@sanitize@url [0]{\catcode `\\12\catcode `\$12\catcode
  `\&12\catcode `\#12\catcode `\^12\catcode `\_12\catcode `\%12\relax}%
\providecommand \@@startlink[1]{}%
\providecommand \@@endlink[0]{}%
\providecommand \url  [0]{\begingroup\@sanitize@url \@url }%
\providecommand \@url [1]{\endgroup\@href {#1}{\urlprefix }}%
\providecommand \urlprefix  [0]{URL }%
\providecommand \Eprint [0]{\href }%
\providecommand \doibase [0]{http://dx.doi.org/}%
\providecommand \selectlanguage [0]{\@gobble}%
\providecommand \bibinfo  [0]{\@secondoftwo}%
\providecommand \bibfield  [0]{\@secondoftwo}%
\providecommand \translation [1]{[#1]}%
\providecommand \BibitemOpen [0]{}%
\providecommand \bibitemStop [0]{}%
\providecommand \bibitemNoStop [0]{.\EOS\space}%
\providecommand \EOS [0]{\spacefactor3000\relax}%
\providecommand \BibitemShut  [1]{\csname bibitem#1\endcsname}%
\let\auto@bib@innerbib\@empty
\bibitem [{\citenamefont {Lindner}\ and\ \citenamefont {Zemb}(1991)}]{ILL1}%
  \BibitemOpen
  \bibfield  {author} {\bibinfo {author} {\bibfnamefont {P.}~\bibnamefont
  {Lindner}}\ and\ \bibinfo {author} {\bibfnamefont {T.}~\bibnamefont {Zemb}},\
  }\href@noop {} {\emph {\bibinfo {title} {Neutron, X-Ray and Light Scattering:
  Introduction to an Investigative Tool for Colloidal and Polymetric
  Systems}}}\ (\bibinfo  {publisher} {North-Holland},\ \bibinfo {address}
  {Amsterdam},\ \bibinfo {year} {1991})\BibitemShut {NoStop}%
\bibitem [{\citenamefont {Lindner}\ and\ \citenamefont {Zemb}(2002)}]{ILL2}%
  \BibitemOpen
  \bibfield  {author} {\bibinfo {author} {\bibfnamefont {P.}~\bibnamefont
  {Lindner}}\ and\ \bibinfo {author} {\bibfnamefont {T.}~\bibnamefont {Zemb}},\
  }\href@noop {} {\emph {\bibinfo {title} {Neutron, X-rays and Light.
  Scattering Methods Applied to Soft Condensed Matter}}},\ \bibinfo {edition}
  {1st}\ ed.\ (\bibinfo  {publisher} {North-Holland},\ \bibinfo {address}
  {Amsterdam},\ \bibinfo {year} {2002})\BibitemShut {NoStop}%
\bibitem [{\citenamefont {Lindner}\ and\ \citenamefont
  {Oberdisse}(2025)}]{ILL3}%
  \BibitemOpen
  \bibfield  {author} {\bibinfo {author} {\bibfnamefont {P.}~\bibnamefont
  {Lindner}}\ and\ \bibinfo {author} {\bibfnamefont {J.}~\bibnamefont
  {Oberdisse}},\ }\href@noop {} {\emph {\bibinfo {title} {Neutron, X-rays and
  Light. Scattering Methods Applied to Soft Condensed Matter}}},\ \bibinfo
  {edition} {2nd}\ ed.\ (\bibinfo  {publisher} {Elsevier},\ \bibinfo {address}
  {Amsterdam},\ \bibinfo {year} {2025})\BibitemShut {NoStop}%
\bibitem [{\citenamefont {F{\"o}rster}\ and\ \citenamefont
  {Antonietti}(1998)}]{forster1998amphiphilic}%
  \BibitemOpen
  \bibfield  {author} {\bibinfo {author} {\bibfnamefont {S.}~\bibnamefont
  {F{\"o}rster}}\ and\ \bibinfo {author} {\bibfnamefont {M.}~\bibnamefont
  {Antonietti}},\ }\href@noop {} {\bibfield  {journal} {\bibinfo  {journal}
  {Adv. Mater.}\ }\textbf {\bibinfo {volume} {10}},\ \bibinfo {pages} {195}
  (\bibinfo {year} {1998})}\BibitemShut {NoStop}%
\bibitem [{\citenamefont {Williams}\ and\ \citenamefont
  {Rasmussen}(1996)}]{williams1996gaussian}%
  \BibitemOpen
  \bibfield  {author} {\bibinfo {author} {\bibfnamefont {C.~K.~I.}\
  \bibnamefont {Williams}}\ and\ \bibinfo {author} {\bibfnamefont {C.~E.}\
  \bibnamefont {Rasmussen}},\ }\href@noop {} {\bibfield  {journal} {\bibinfo
  {journal} {Adv. Neural Inf. Process. Syst.}\ }\textbf {\bibinfo {volume}
  {8}},\ \bibinfo {pages} {514} (\bibinfo {year} {1996})}\BibitemShut {NoStop}%
\bibitem [{\citenamefont {Seeger}(2004)}]{seeger2004gaussian}%
  \BibitemOpen
  \bibfield  {author} {\bibinfo {author} {\bibfnamefont {M.}~\bibnamefont
  {Seeger}},\ }\href@noop {} {\bibfield  {journal} {\bibinfo  {journal} {Int.
  J. Neural Syst.}\ }\textbf {\bibinfo {volume} {14}},\ \bibinfo {pages} {69}
  (\bibinfo {year} {2004})}\BibitemShut {NoStop}%
\bibitem [{\citenamefont {Rasmussen}\ and\ \citenamefont
  {Williams}(2006)}]{GPR}%
  \BibitemOpen
  \bibfield  {author} {\bibinfo {author} {\bibfnamefont {C.~E.}\ \bibnamefont
  {Rasmussen}}\ and\ \bibinfo {author} {\bibfnamefont {C.~K.}\ \bibnamefont
  {Williams}},\ }\href@noop {} {\emph {\bibinfo {title} {Gaussian Processes for
  Machine Learning}}}\ (\bibinfo  {publisher} {MIT Press},\ \bibinfo {address}
  {Cambridge},\ \bibinfo {year} {2006})\BibitemShut {NoStop}%
\bibitem [{\citenamefont {Deringer}\ \emph {et~al.}(2021)\citenamefont
  {Deringer}, \citenamefont {Bartók}, \citenamefont {Bernstein}, \citenamefont
  {Wilkins}, \citenamefont {Ceriotti},\ and\ \citenamefont
  {Csányi}}]{Deringer2021Gaussian}%
  \BibitemOpen
  \bibfield  {author} {\bibinfo {author} {\bibfnamefont {V.~L.}\ \bibnamefont
  {Deringer}}, \bibinfo {author} {\bibfnamefont {A.~P.}\ \bibnamefont
  {Bartók}}, \bibinfo {author} {\bibfnamefont {N.}~\bibnamefont {Bernstein}},
  \bibinfo {author} {\bibfnamefont {D.~M.}\ \bibnamefont {Wilkins}}, \bibinfo
  {author} {\bibfnamefont {M.}~\bibnamefont {Ceriotti}}, \ and\ \bibinfo
  {author} {\bibfnamefont {G.}~\bibnamefont {Csányi}},\ }\href@noop {}
  {\bibfield  {journal} {\bibinfo  {journal} {Chem. Rev.}\ }\textbf {\bibinfo
  {volume} {121}},\ \bibinfo {pages} {10073} (\bibinfo {year}
  {2021})}\BibitemShut {NoStop}%
\bibitem [{\citenamefont {Tung}\ \emph {et~al.}(2025)\citenamefont {Tung},
  \citenamefont {Yip}, \citenamefont {Huang}, \citenamefont {Porcar},
  \citenamefont {Shinohara}, \citenamefont {Sumpter}, \citenamefont {Ding},
  \citenamefont {Do},\ and\ \citenamefont {Chen}}]{Tung2025}%
  \BibitemOpen
  \bibfield  {author} {\bibinfo {author} {\bibfnamefont {C.-H.}\ \bibnamefont
  {Tung}}, \bibinfo {author} {\bibfnamefont {S.}~\bibnamefont {Yip}}, \bibinfo
  {author} {\bibfnamefont {G.-R.}\ \bibnamefont {Huang}}, \bibinfo {author}
  {\bibfnamefont {L.}~\bibnamefont {Porcar}}, \bibinfo {author} {\bibfnamefont
  {Y.}~\bibnamefont {Shinohara}}, \bibinfo {author} {\bibfnamefont {B.~G.}\
  \bibnamefont {Sumpter}}, \bibinfo {author} {\bibfnamefont {L.}~\bibnamefont
  {Ding}}, \bibinfo {author} {\bibfnamefont {C.}~\bibnamefont {Do}}, \ and\
  \bibinfo {author} {\bibfnamefont {W.-R.}\ \bibnamefont {Chen}},\ }\href@noop
  {} {\bibfield  {journal} {\bibinfo  {journal} {J. Colloid Interface Sci.}\
  }\textbf {\bibinfo {volume} {692}},\ \bibinfo {pages} {137554} (\bibinfo
  {year} {2025})}\BibitemShut {NoStop}%
\bibitem [{\citenamefont {Jeffreys}(1984)}]{jeffreys1939}%
  \BibitemOpen
  \bibfield  {author} {\bibinfo {author} {\bibfnamefont {H.}~\bibnamefont
  {Jeffreys}},\ }\href@noop {} {\emph {\bibinfo {title} {Theory of
  Probability}}},\ \bibinfo {edition} {3rd}\ ed.\ (\bibinfo  {publisher}
  {Oxford University Press},\ \bibinfo {address} {Oxford},\ \bibinfo {year}
  {1984})\BibitemShut {NoStop}%
\bibitem [{\citenamefont {Gelman}\ \emph {et~al.}(2013)\citenamefont {Gelman},
  \citenamefont {Carlin}, \citenamefont {Stern}, \citenamefont {Dunson},
  \citenamefont {Vehtari},\ and\ \citenamefont {Rubin}}]{Gelman2013}%
  \BibitemOpen
  \bibfield  {author} {\bibinfo {author} {\bibfnamefont {A.}~\bibnamefont
  {Gelman}}, \bibinfo {author} {\bibfnamefont {J.~B.}\ \bibnamefont {Carlin}},
  \bibinfo {author} {\bibfnamefont {H.~S.}\ \bibnamefont {Stern}}, \bibinfo
  {author} {\bibfnamefont {D.~B.}\ \bibnamefont {Dunson}}, \bibinfo {author}
  {\bibfnamefont {A.}~\bibnamefont {Vehtari}}, \ and\ \bibinfo {author}
  {\bibfnamefont {D.~B.}\ \bibnamefont {Rubin}},\ }\href@noop {} {\emph
  {\bibinfo {title} {Bayesian Data Analysis}}},\ \bibinfo {edition} {3rd}\ ed.\
  (\bibinfo  {publisher} {CRC Press},\ \bibinfo {address} {Boca Raton},\
  \bibinfo {year} {2013})\BibitemShut {NoStop}%
\bibitem [{\citenamefont {Lam}\ \emph {et~al.}(2019)\citenamefont {Lam},
  \citenamefont {Do}, \citenamefont {Wang}, \citenamefont {Huang},\ and\
  \citenamefont {Chen}}]{lam2019ctab}%
  \BibitemOpen
  \bibfield  {author} {\bibinfo {author} {\bibfnamefont {C.~N.}\ \bibnamefont
  {Lam}}, \bibinfo {author} {\bibfnamefont {C.}~\bibnamefont {Do}}, \bibinfo
  {author} {\bibfnamefont {Y.}~\bibnamefont {Wang}}, \bibinfo {author}
  {\bibfnamefont {G.-R.}\ \bibnamefont {Huang}}, \ and\ \bibinfo {author}
  {\bibfnamefont {W.-R.}\ \bibnamefont {Chen}},\ }\href@noop {} {\bibfield
  {journal} {\bibinfo  {journal} {Phys. Chem. Chem. Phys.}\ }\textbf {\bibinfo
  {volume} {21}},\ \bibinfo {pages} {18346} (\bibinfo {year}
  {2019})}\BibitemShut {NoStop}%
\bibitem [{\citenamefont {Huang}\ \emph {et~al.}(2022)\citenamefont {Huang},
  \citenamefont {Lam}, \citenamefont {Hong}, \citenamefont {Wang},
  \citenamefont {Shinohara}, \citenamefont {Do},\ and\ \citenamefont
  {Chen}}]{huang2022ion}%
  \BibitemOpen
  \bibfield  {author} {\bibinfo {author} {\bibfnamefont {G.-R.}\ \bibnamefont
  {Huang}}, \bibinfo {author} {\bibfnamefont {C.~N.}\ \bibnamefont {Lam}},
  \bibinfo {author} {\bibfnamefont {K.}~\bibnamefont {Hong}}, \bibinfo {author}
  {\bibfnamefont {Y.}~\bibnamefont {Wang}}, \bibinfo {author} {\bibfnamefont
  {Y.}~\bibnamefont {Shinohara}}, \bibinfo {author} {\bibfnamefont
  {C.}~\bibnamefont {Do}}, \ and\ \bibinfo {author} {\bibfnamefont {W.-R.}\
  \bibnamefont {Chen}},\ }\href@noop {} {\bibfield  {journal} {\bibinfo
  {journal} {ACS Macro Lett.}\ }\textbf {\bibinfo {volume} {11}},\ \bibinfo
  {pages} {66} (\bibinfo {year} {2022})}\BibitemShut {NoStop}%
\bibitem [{\citenamefont {James}(2006)}]{james2006statistical}%
  \BibitemOpen
  \bibfield  {author} {\bibinfo {author} {\bibfnamefont {F.}~\bibnamefont
  {James}},\ }\href@noop {} {\emph {\bibinfo {title} {Statistical Methods in
  Experimental Physics}}},\ \bibinfo {edition} {2nd}\ ed.\ (\bibinfo
  {publisher} {World Scientific},\ \bibinfo {address} {Singapore},\ \bibinfo
  {year} {2006})\BibitemShut {NoStop}%
\bibitem [{\citenamefont {Huang}(1991)}]{huang1991statistical}%
  \BibitemOpen
  \bibfield  {author} {\bibinfo {author} {\bibfnamefont {K.}~\bibnamefont
  {Huang}},\ }\href@noop {} {\emph {\bibinfo {title} {Statistical
  Mechanics}}},\ \bibinfo {edition} {2nd}\ ed.\ (\bibinfo  {publisher} {John
  Wiley \& Sons},\ \bibinfo {address} {New York},\ \bibinfo {year}
  {1991})\BibitemShut {NoStop}%
\bibitem [{\citenamefont {Kardar}(2007)}]{kardar2007statistical}%
  \BibitemOpen
  \bibfield  {author} {\bibinfo {author} {\bibfnamefont {M.}~\bibnamefont
  {Kardar}},\ }\href@noop {} {\emph {\bibinfo {title} {Statistical Physics of
  Particles}}}\ (\bibinfo  {publisher} {Cambridge University Press},\ \bibinfo
  {address} {Cambridge},\ \bibinfo {year} {2007})\BibitemShut {NoStop}%
\bibitem [{\citenamefont {Anderson}\ \emph {et~al.}(2016)\citenamefont
  {Anderson}, \citenamefont {Andreani}, \citenamefont {Carpenter},
  \citenamefont {Festa}, \citenamefont {Gorini}, \citenamefont {Loong},\ and\
  \citenamefont {Senesi}}]{carpenter2016research}%
  \BibitemOpen
  \bibfield  {author} {\bibinfo {author} {\bibfnamefont {I.}~\bibnamefont
  {Anderson}}, \bibinfo {author} {\bibfnamefont {C.}~\bibnamefont {Andreani}},
  \bibinfo {author} {\bibfnamefont {J.}~\bibnamefont {Carpenter}}, \bibinfo
  {author} {\bibfnamefont {G.}~\bibnamefont {Festa}}, \bibinfo {author}
  {\bibfnamefont {G.}~\bibnamefont {Gorini}}, \bibinfo {author} {\bibfnamefont
  {C.-K.}\ \bibnamefont {Loong}}, \ and\ \bibinfo {author} {\bibfnamefont
  {R.}~\bibnamefont {Senesi}},\ }\href@noop {} {\bibfield  {journal} {\bibinfo
  {journal} {Phys. Rep.}\ }\textbf {\bibinfo {volume} {654}},\ \bibinfo {pages}
  {1} (\bibinfo {year} {2016})}\BibitemShut {NoStop}%
\bibitem [{\citenamefont {Brückel}\ \emph {et~al.}(2020)\citenamefont
  {Brückel}, \citenamefont {Gutberlet}, \citenamefont {Schmidt}, \citenamefont
  {Alba-Simionesco}, \citenamefont {Ott},\ and\ \citenamefont
  {Menelle}}]{Bruckel2020}%
  \BibitemOpen
  \bibfield  {author} {\bibinfo {author} {\bibfnamefont {T.}~\bibnamefont
  {Brückel}}, \bibinfo {author} {\bibfnamefont {T.}~\bibnamefont {Gutberlet}},
  \bibinfo {author} {\bibfnamefont {S.}~\bibnamefont {Schmidt}}, \bibinfo
  {author} {\bibfnamefont {C.}~\bibnamefont {Alba-Simionesco}}, \bibinfo
  {author} {\bibfnamefont {F.}~\bibnamefont {Ott}}, \ and\ \bibinfo {author}
  {\bibfnamefont {A.}~\bibnamefont {Menelle}},\ }\href@noop {} {\bibfield
  {journal} {\bibinfo  {journal} {Neutron News}\ }\textbf {\bibinfo {volume}
  {31}},\ \bibinfo {pages} {14} (\bibinfo {year} {2020})}\BibitemShut {NoStop}%
\bibitem [{IAE(2021)}]{IAEATECDOC1981}%
  \BibitemOpen
  \href@noop {} {\emph {\bibinfo {title} {Compact Accelerator Based Neutron
  Sources}}},\ \bibinfo {type} {Tech. Rep.}\ \bibinfo {number} {1981}\
  (\bibinfo  {institution} {International Atomic Energy Agency},\ \bibinfo
  {address} {Vienna},\ \bibinfo {year} {2021})\BibitemShut {NoStop}%
\bibitem [{\citenamefont {Ott}\ \emph {et~al.}(2023)\citenamefont {Ott},
  \citenamefont {Darpentigny}, \citenamefont {Annighöfer}, \citenamefont
  {Paulin}, \citenamefont {Meuriot}, \citenamefont {Menelle}, \citenamefont
  {Sellami},\ and\ \citenamefont {Schwindling}}]{ott2023icone}%
  \BibitemOpen
  \bibfield  {author} {\bibinfo {author} {\bibfnamefont {F.}~\bibnamefont
  {Ott}}, \bibinfo {author} {\bibfnamefont {J.}~\bibnamefont {Darpentigny}},
  \bibinfo {author} {\bibfnamefont {B.}~\bibnamefont {Annighöfer}}, \bibinfo
  {author} {\bibfnamefont {M.~A.}\ \bibnamefont {Paulin}}, \bibinfo {author}
  {\bibfnamefont {J.-L.}\ \bibnamefont {Meuriot}}, \bibinfo {author}
  {\bibfnamefont {A.}~\bibnamefont {Menelle}}, \bibinfo {author} {\bibfnamefont
  {N.}~\bibnamefont {Sellami}}, \ and\ \bibinfo {author} {\bibfnamefont
  {J.}~\bibnamefont {Schwindling}},\ }\href@noop {} {\bibfield  {journal}
  {\bibinfo  {journal} {EPJ Web Conf.}\ }\textbf {\bibinfo {volume} {286}},\
  \bibinfo {pages} {02001} (\bibinfo {year} {2023})}\BibitemShut {NoStop}%
\end{thebibliography}

%

\end{document}